\documentclass[prl, reprint, amsmath, amssymb, aps, nofootinbib, nobibnotes, superscriptaddress]{revtex4-2}
\usepackage{graphicx}
\usepackage{hyperref}
\usepackage{dcolumn}
\usepackage{bm}
\usepackage{color}
\usepackage{amsmath}
\usepackage[x11names]{xcolor}
\usepackage{orcidlink}

\providecommand{\prd}{Phys. Rev. D}

\providecommand{\apj}{Astrophys. J.}
\providecommand{\mnras}{Mon. Not. R. Astron. Soc.}
\providecommand{\aap}{Astron. Astrophys.}
\providecommand{\jcap}{JCAP}

\begin{document}
\title{Breaking the background and linear degeneracies: \\Using nonlinear clustering to discriminate between dark energy models}
\author{Yunhao Gao~\orcidlink{0009-0000-6958-1139}}\email{gaoyh@nao.cas.cn}
\affiliation{National Astronomical Observatories, Chinese Academy of Sciences, Beijing, 100101, China}
\affiliation{School of Astronomy and Space Sciences, University of Chinese Academy of Sciences, Beijing, 100049, China}
\affiliation{Institute for Computational Cosmology, Department of Physics, Durham University, Durham DH1 3LE, UK}

\author{Baojiu Li~\orcidlink{0000-0002-1098-9188}}\email{baojiu.li@durham.ac.uk}
\affiliation{Institute for Computational Cosmology, Department of Physics, Durham University, Durham DH1 3LE, UK}
\affiliation{National Astronomical Observatories, Chinese Academy of Sciences, Beijing, 100101, China}

\author{Jie Wang~\orcidlink{0000-0002-9937-2351}}\email{jie.wang@nao.cas.cn}
\affiliation{National Astronomical Observatories, Chinese Academy of Sciences, Beijing, 100101, China}
\affiliation{School of Astronomy and Space Sciences, University of Chinese Academy of Sciences, Beijing, 100049, China}
\affiliation{\mbox{Institute for Frontiers in Astronomy and Astrophysics, Beijing Normal University, Beijing, 102206, China}}

\author{Carlton M. Baugh~\orcidlink{0000-0002-9935-9755}}\email{c.m.baugh@durham.ac.uk}
\affiliation{Institute for Computational Cosmology, Department of Physics, Durham University, Durham DH1 3LE, UK}

\begin{abstract}
    We show that the traditional approach of parameterizing the background and linear-perturbation evolutions, through the equation-of-state parameter $w(a)$ and a time-dependent sound speed squared $\bar{c}^2_s(a)$, does not fully specify the cosmological behavior of dark energy theories. Using a new, fully nonlinear fluid framework, we perform spherical-collapse simulations of two clustering dark energy models with identical $w(a)$ and $\bar{c}^2_s(a)$, finding that their central dark-energy densities differ by up to 50\% for cluster-sized dark matter halos. We identify the departure of the local sound speed from $\bar{c}_s$ as the main driver of these differences. In contrast, all other theory-specific quantities in the fluid equations can be described by their background values to a very good approximation. The results demonstrate that nonlinear structure formation encodes critical additional information about the underlying dark energy theory, opening an avenue for discriminating between models using observables on small scales.
\end{abstract}

\maketitle

\textbf{Introduction -}
Can cosmological observations distinguish dark-energy (DE) theories that share the same background evolution and linear perturbation dynamics? The properties of DE are commonly parameterized by its equation of state $w(a)$ and effective sound speed squared $\bar{c}_s^2(a)$, characterizing its evolution on the background and linear scales. However, these quantities do not uniquely determine the underlying scalar-field theory \cite[e.g.,][]{Degeneracy}, raising the possibility that models indistinguishable on the background and linear levels may exhibit different nonlinear behavior. For clustering DE, a major class of DE models, this has so far remained an important but open question that awaits a quantitative answer, since for that we must be able to predict the evolution of these models well into the nonlinear regime, a topic we focus on here.

Recent results from DESI have renewed interest in dynamical DE models \cite{DESI01, DESI02, DESI03}, while other large surveys such as Euclid \cite{Euclid01, Euclid02, Euclid03} and LSST \cite{LSST01, LSST02} aim to probe the growth of cosmic structure with unprecedented precision. Fully exploiting data of these observations to constrain the underlying DE theories requires accurate predictions of their effects on structure formation, particularly in the nonlinear regime, which is a big challenge. Considerable efforts have been devoted to this problem. Spherical collapse has been used to investigate the impact of clustering DE on halo formation and abundance \cite{SC01, SC02, SC03, SC04, SC05, SC06}, and perturbative approaches explored to quantify its effects on the growth of large-scale structure \cite{Perturbation_theory01, Perturbation_theory02, Perturbation_theory03}. More recently, cosmological simulations have enabled the study of DE clustering in nonlinear environments \cite{k-evolution01, k-evolution02, FVM01, FVM02}, including its impact on halo properties and gravitational 
observables such as weak lensing and the integrated Sachs-Wolfe effect \cite{k-evolution-observation}.

Despite these progresses, existing numerical implementations commonly employ effective parameterizations or small-quantity expansions for the underlying field dynamics, rather than evolving the complete nonlinear dynamics of the scalar field. Such approximations may not fully capture certain theory-dependent effects that become important in the strongly nonlinear regime.

In this Letter, we present a fully nonlinear and self-consistent fluid framework that enables the evolution of clustering DE to be simulated from its underlying scalar-field Lagrangian, based on an action-splitting method detailed in an accompanying paper \cite{CompanionPRD}. Unlike approaches relying on a prescribed background equation of state and sound speed, this formulation preserves the dependence of the fluid dynamics on the underlying theory, without requiring a perturbative expansion of the scalar-field or fluid variables. Applying it to spherical-collapse simulations, we demonstrate that DE models sharing the same background and linear dynamics can develop significantly different nonlinear density profiles. We show that these differences are mainly due to deviations of the local sound speed from its background value, while the role of other theory-dependent quantities in the evolution equations is negligible. 

\textbf{The Fluid Framework -}
We consider the $k$-essence model whose action can be written as
\begin{equation}\label{k-essence action}
    S=\int \text{d}^4x\sqrt{-g}K(\phi, X),
\end{equation}
where $g$ is the determinant of the metric $g_{ab}$ and $X=g^{ab}\partial_a\phi\partial_b\phi$. We define the 4-velocity of DE to be $u^a=-\nabla^a\phi/\sqrt{-X}$ and split the Lagrangian $K$ into two parts: the kinetic part $XK_X$ whose density $\rho_k$ and pressure $P_k$ satisfy $P_k=\rho_k=XK_X$, and the potential part with the relation $P_V=-\rho_V=K-XK_X$. The equations of state for the kinetic and potential parts hold on all scales, not just on the background.

\begin{figure*}[t]
\includegraphics[width=\textwidth]{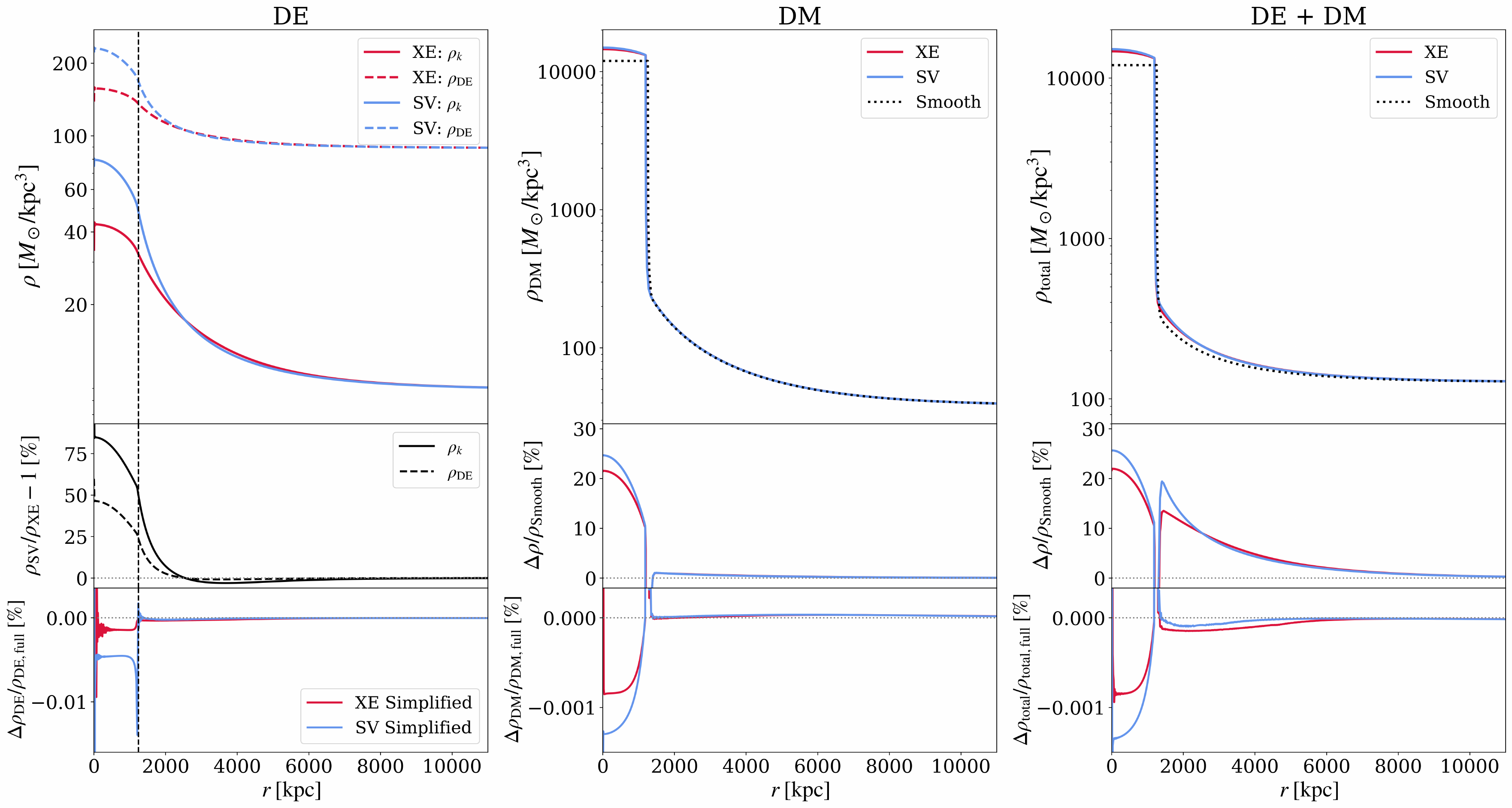}
\caption{The density profiles of dark energy, dark matter, and their sum at $z=0$. {\textit{First Column:}} The upper panel presents the profiles of the kinetic density $\rho_k=XK_X$ (solid lines) and the total DE energy density $\rho_{\rm DE}=2XK_X-K$ (dashed lines). The middle panel presents the relative difference of these densities between the SV and XE models. The lower panel presents the relative difference in the total DE density between the simplified and full simulations (see the text for more details), for both models. Red and blue colors indicate the XE and SV models respectively, a color scheme used in this whole figure. The vertical dashed line marks the average radius of the DM halo in the two models. {\textit{Second Column:}} The upper and lower panels are the same as in the first column, but for the DM density profile. In the upper panel the dotted line shows the DM profile assuming smooth (non-clustering) DE. The middle panel shows the relative difference of the DM density profile between clustering and smooth DE. 
The sharp feature at $r\simeq1$--$1.5\;{\rm Mpc}$ in the lower panel is caused by the slight mismatch in the halo radii of these models. 
{\textit{Third Column:}} Same as the second column, but for the total density $\rho_{\rm total}=\rho_{\rm DM} + \rho_{\rm DE}$.
}
\label{fig:density profiles}
\end{figure*}

We write the conservation equation $\nabla_bT^{ab}_{\rm DE}=0$, where $T^{ab}_{\rm DE}$ is the DE energy-momentum tensor, 
into a set of fluid equations for the \textit{kinetic} part of the DE: 
\begin{equation}\label{fluid equations}
\begin{aligned}
    \frac{\text{d}\textcolor{Coral2}{\rho_k}}{\text{d}\tau}+\left[\textcolor{Coral2}{\rho_k}+\textcolor{Coral2}{P_k}c_s^2\right] \nabla_a\textcolor{RoyalBlue1}{u^a}&=S(\textcolor{Green4}{\phi}, X),\\
    \left[\textcolor{Coral2}{\rho_k}+\textcolor{Coral2}{P_k}c_s^2\right]\frac{\text{D}\textcolor{RoyalBlue1}{u^a}}{\text{d}\tau}&=-c_s^2\left(g^{ab}+\textcolor{RoyalBlue1}{u^a}\textcolor{RoyalBlue1}{u^b}\right)\partial_b\textcolor{Coral2}{P_k},\\
    \frac{\text{d}\textcolor{Green4}{\phi}}{\text{d}\tau}&=\sqrt{-X\left(\textcolor{Coral2}{\rho_k}, \textcolor{Green4}{\phi}\right)},
\end{aligned}
\end{equation}
with $S$ in the continuity equation being the source terms
\begin{equation}\label{source terms}
    S=\frac{1+c_s^2}{2}\sqrt{-X}K_{\textcolor{Green4}{\phi}} + c_s^2\left(-X\right)^{3/2}K_{\textcolor{Green4}{\phi}X},
\end{equation}
and $c_s^2=K_X/\left(K_X+2XK_{XX}\right)$ is the local sound speed squared. We invert the relation $\rho_k=XK_X$ to obtain the value of $X$. The scalar field is assumed to be minimally coupled to gravity, meaning that it interacts with matter only through gravitational force. In the Newtonian gauge
\begin{equation}
    \text{d}s^2=-\left(1+2\Psi\right)\text{d}t^2+a^2\left(1-2\Phi\right)\delta_{ij}\text{d}x^i\text{d}x^j,
\end{equation}
and on small scales, the Poisson equation becomes
\begin{equation}\label{Poisson equation}
    \nabla^2\Psi=4\pi Ga^2(\delta\rho_m+2\delta\rho_k+2\delta K(\phi, X)),
\end{equation}
where $a$ is the scale factor. 

Deriving Eqs.~\eqref{fluid equations} requires no small-quantity expansion: one can show that, after substituting the definition of $\rho_k$ and $u^a$ in terms of the scalar field, the continuity equation is equivalent to the equation of motion 
of the $k$-essence theory, while the other two equations are pure identities. These equations are closed and self-consistent. A detailed derivation of them, together with their specific forms in spherical collapse simulations, can be found in \cite{CompanionPRD}.

\textbf{Breaking the model degeneracy in the nonlinear regime -}
We apply Eqs.~\eqref{fluid equations} and~\eqref{Poisson equation} to a set of spherical collapse simulations with dark matter (DM) and DE components. Two $k$-essence models are considered, respectively what we call XE and SV, defined as
\begin{equation}\label{XE and SV}
\begin{aligned}
    K_{\rm XE}&=V(\phi)+f(\phi)X+q(\phi)X^2,\\
    K_{\rm SV}&=q(\phi)h(X).
\end{aligned}
\end{equation}
These functions are uniquely reconstructed \cite{Reconstruction} for a background expansion history of $w_0=-0.8, \,w_a=-0.1$
(with $w(a)=w_0+w_a(1-a)$), and $\bar{c}_s^2=10^{-6}$. We also fix the density parameters at $z=0$ to 
$\Omega_m=0.3, \,\Omega_{\rm DE}=0.7$ and the Hubble constant to $H_0=67.66{\rm \;km\cdot s^{-1}\cdot Mpc^{-1}}$. Consequently, these models 
have identical background and linear dynamics.

The initial condition of DM is set to be the following top-hat profile at $z=49$,
\begin{equation}\label{initial tophat}
    \delta_{\rm DM, ini}(r)=\left\{
    \begin{array}{cc}
        0.076 & \quad r\leq R_{\rm halo,ini} \\
        0 & \quad r > R_{\rm halo,ini},
    \end{array}
    \right.
\end{equation}
where the halo mass $M_{\rm DM}(<R_{\rm halo,ini})=10^{14}M_\odot$, while the initial DE density perturbation is assumed to be zero at this redshift. 
We then use Eq.~\eqref{fluid equations} and~\eqref{Poisson equation} to numerically evolve this system to $z=0$, when the physical radius of the DM halo collapses to around half of its turnaround radius. The density profiles for DE, DM, and DE + DM are plotted in the first two rows of Figure~\ref{fig:density profiles}.

Despite the degeneracy in the background and linear evolution, the XE and SV models develop distinct density distributions in the central region of the DM halo, where non-linearity reaches its maximum. The kinetic density and the total DE density of the SV model are around 80\% and 50\% higher than those of the XE model in these regions. Though slightly less significant, these differences also affect the dark-matter distribution through the pure gravitational force, leading to a few-percent discrepancy in the total DM + DE density profiles. This discrepancy, 
which is relevant to observations, is still appreciable up to around 2 Mpc. This result reveals that the nonlinear evolution of clustering DE indeed retains information about the underlying scalar-field theory that is not encoded in the background expansion history or sound speed.

\begin{figure*}[t]
\includegraphics[width=\textwidth]{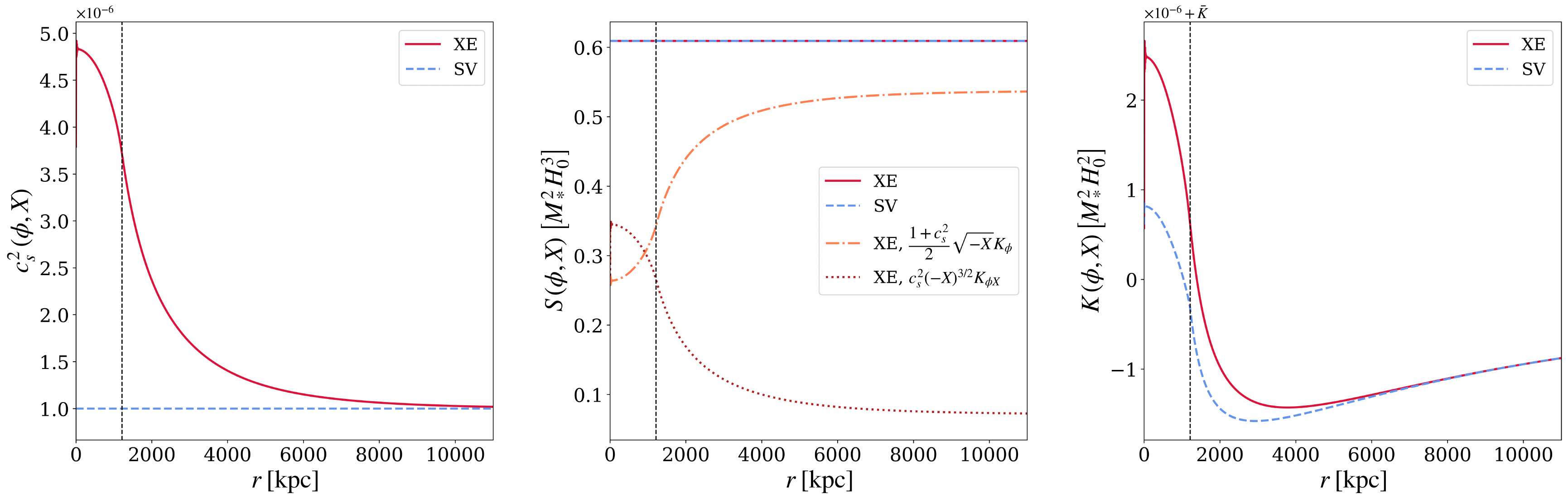}
\caption{Profiles of the sound speed squared $c_s^2(\phi, X)$ (first panel), the source terms $S(\phi, X)$ (second panel; normalized by $M_*^2H_0^3$; see Eq.~\eqref{source terms} for the definition), 
and $K(\phi, X)$ (third panel; with $\bar{K}\approx-1.68M_*^2H_0^2$ and normalized by $M_*^2H_0^2$). We present the profiles for the XE model (red solid line) and the SV model (blue dashed line) at $z=0$, and 
separately plot the two terms within $S$ for the XE model in the middle panel. Note that we also mark the average physical DM-halo radius for XE and SV models at $z=0$ (vertical dotted lines).
}
\label{fig:SoundSpeed_and_Source}
\end{figure*}

\textbf{Origin of the Nonlinear Difference -}
To understand the origin of this phenomenon, we examine the physical quantities in Eqs.~(\ref{fluid equations}, \ref{Poisson equation}) that are model-dependent: $c_s^2(\phi, X)$, 
$S(\phi, X)$, and $K(\phi, X)$. A comparison of the profiles of these quantities at $z=0$ for XE and SV models is displayed in Figure~\ref{fig:SoundSpeed_and_Source}.

The first panel of Figure~\ref{fig:SoundSpeed_and_Source} shows that 
the XE and SV models lead to different local values of $c_s^2$. In particular, the increase of $c_s^2$ in the XE model in the central region strengthens the influence of the $c_s^2\partial P_k$ term in the second equation of Eq.~\eqref{fluid equations}, and thus enhances the suppression of DE clustering. On the other hand, the source terms and $K(\phi, X)$ are almost radius-independent and model-independent. Indeed, for the XE model, the two terms within 
$S$, cf.~Eq.~\eqref{source terms}, are also plotted in the middle panel: while they vary with radius, 
their sum does not. We checked the 
simulation outputs and found that, over the simulated period, the relative deviation of $S$ or $K$ from their background value is of order $\mathcal{O}(10^{-6})$.

It can be shown that these results are general for clustering $k$-essence theories by analyzing the linear response of these quantities to the perturbation of the density and the scalar field. We present the key results here, leaving the detailed derivations to the appendix.

First, for $c_s^2$, we define an auxiliary function $F(\phi, X)=XK_{XX}/K_X$, so that $c_s^2=1/(1+2F)$, and perturbations in $c_s^2$ can be attributed to perturbations of $F$. Using the field redefinition to make $\bar{\phi}=a$ and thus $\bar{X}=-a^2H^2$ on the background, one can prove that 
\begin{equation}\label{d ln F}
\begin{aligned}
    \delta\ln{|F|}=&\,\frac{1}{1+w}\left(2\bar{c}_s^2Q-1\right)\delta\ln{\rho}_{\rm DE}\\&+\left(6\bar{c}_s^2Q-3-\frac{(\ln{\bar{c}_s^2})'}{1-\bar{c}_s^2}\right)\delta\ln{|\phi|},
\end{aligned}
\end{equation}
where $Q=\dfrac{XK_{XXX}}{K_{XX}}+\dfrac{3}{2}$ and $(\ln{c_s^2})'=\dfrac{\text{d}\ln{\bar{c}_s^2}}{\text{d}\ln{a}}$. The coefficients of $\delta\ln{\rho}_{\rm DE}$ and $\delta\ln{|\phi|}$ are evaluated on the background, because we are considering perturbations 
with respect to it. For clustering DE theories, we have $\bar{c}_s^2\ll1$ and $\delta\ln{|\phi|}=\delta\phi/\bar{\phi}=\mathcal{O}(10^{-6})$, which have been checked directly in simulations. If we further assume $Q$ is at most $\mathcal{O}(1/\bar{c}_s^2)$ and the rate of change $(\ln{\bar{c}_s^2})'$ is at most $\mathcal{O}(1)$ \footnote{Though these assumptions are not satisfied by all DE models, they are still generic enough for most of the $k$-essence models we are interested in. In addition, even if $Q$ and $(\ln{\bar{c}_s^2})'$ break these assumptions, the following argument for $c_s^2$ is still correct. However, according to Eq.~\eqref{d S}, the value of the source terms can deviate from the background value in that case.}, then the pre-factor of $\delta\ln{|\phi|}$ will be at most $\mathcal{O}(1)$. Consequently, the second line of Eq.~\eqref{d ln F} only contributes to the perturbation of $F$ by at most $\mathcal{O}(10^{-6})$. However, in the first line, $\delta\ln{\rho_{\rm DE}}=\delta\rho_{\rm DE}/\bar{\rho}_{\rm DE}$ can become $\mathcal{O}(1)$ at non-linear scales, so if the pre-factor of this term is also 
$\mathcal{O}(1)$,it can lead to considerable perturbations in $F$ and thereby $c_s^2$. This implies that it is the nonlinear perturbation of the DE density that leads to the deviation of local $c_s^2$ from its background value.

Second, for $K(\phi, X)$, one can prove that
\begin{equation}\label{d ln K}
    \delta\ln{|K|}=\frac{\bar{c}_s^2}{w}\delta\ln{\rho}_{\rm DE}+\frac{3(1+w)}{w}(\bar{c}_s^2-c_a^2)\delta\ln{|\phi|},
\end{equation}
where $c^2_a(a)=\dfrac{\dot{P}_{\rm DE}}{\dot{\rho}_{\rm DE}}=w(a)-\dfrac{w'(a)}{3(1+w(a))}$ is the adiabatic sound speed squared on the background. Provided that $w$ stays sufficiently far from the phantom divide ($w=-1$), $c_a^2$ remains of $\mathcal{O}(1)$, and so does the coefficient of $\delta\ln{|\phi|}$, so the contribution of the second term to the $K$ perturbation is $\mathcal{O}(10^{-6})$. Moreover, in this case, the prefactor of $\delta\ln{\rho}_{\rm DE}$ is 
$\mathcal{O}(\bar{c}^2_s)\ll1$, so the first term also contributes negligibly to the $K$ perturbation. Therefore, $|\delta K|\ll|K|$ and 
$\delta K$ can be neglected in simulations. 

Finally, for source terms $S(\phi, X)$, we have
\begin{equation}\label{d S}
    \delta S = \frac{2\bar{c}_s^2}{1+w}H\rho_k T_1(a)\delta\ln{\rho}_{\rm DE}+H\rho_k T_2(a)\delta\ln{|\phi|}.
\end{equation}
Here we do not consider $\delta\ln{|S|}$ because $\bar{S}$ can vanish on the background when $c_a^2(a)=\bar{c}_s^2(a)$. The $T_i$ functions can be computed as (on the background):
\begin{equation}
\begin{aligned}
    T_1=&\,\frac{w'}{2(1+w)}+\frac{1}{2}(\ln \bar{c}_s^2)'-\frac{3}{2}(w-1)-3\bar{c}_s^2(1-\bar{c}_s^2)Q,\\
    T_2=&\,\frac{w''}{1+w}+3(\bar{c}_s^2)'-6w'-\frac{9}{2}(3+2w+w\Omega_{\rm DE})(\bar{c}_s^2-w)\\&+\frac{3w'}{1+w}\left(\bar{c}_s^2-w-\frac{1}{2}(1+w\Omega_{\rm DE})\right)+6\bar{c}_s^2T_1.
\end{aligned}
\end{equation}
Here, $\Omega_{\rm DE}(a)$ is the density parameter of DE at $a$. As before, if we assume $w'$, $w''$, $(\ln{\bar{c}_s^2})'$ to be at most $\mathcal{O}(1)$, $Q$ to be at most $\mathcal{O}(1/\bar{c}_s^2)$, and $w$ to stay sufficiently far from the phantom divide, all terms in $T_1$ and $T_2$ are at most $\mathcal{O}(1)$. In this case, from the first equation of Eq.~\eqref{fluid equations}, the contribution of $\delta S$ to the evolution of the local $\rho_k$ is 
\begin{equation}
\begin{aligned}
    \frac{\text{d}\ln{\rho_k}}{\text{d}\ln{a}}\Bigg|_{{\rm from\;} \delta S}&=\frac{1}{H\rho_k}\delta S\sim\mathcal{O}(\bar{c}_s^2)+\mathcal{O}(\delta\ln{|\phi|})\ll1,
\end{aligned}
\end{equation}
which can therefore be neglected on the non-linear scales.

Therefore, the physical origin of this nonlinear degeneracy breaking can be traced primarily to the local sound speed squared, whose evolution depends on the underlying scalar-field theory. In contrast, the perturbations of $S(\phi, X)$ and $K(\phi, X)$ remain subdominant at nonlinear scales, as long as the theory satisfies (1) $w(a)\not\to-1$, (2) the EoS and sound speed squared do not change too fast, namely $w'$, $w''$, and $(\ln{\bar{c}_s^2})'$ are at most of order $\mathcal{O}(1)$, (3) 
$Q={XK_{XXX}}/{K_{XX}}+{3}/{2}$ is at most $\mathcal{O}(1/\bar{c}_s^2)$.


We repeat the simulations with $\phi$, $K$ and $S$ fixed to their background values, referred to as the simplified simulations. As shown in the bottom row of Fig.~\ref{fig:density profiles}, the relative differences from the 
full simulations are $\lesssim 10^{-4}$ for DE and $\lesssim 10^{-5}$ for DM and the total density. This confirms the sound speed squared as the only quantity of the underlying scalar-field theory that must be retained to accurately describe its nonlinear evolution. 

\textbf{Implications and Outlook -} 
Our results suggest that the nonlinear regime offers a new opportunity to distinguish dark-energy theories that are degenerate at the background and linear levels. The theory dependence of the local sound speed provides a concrete physical mechanism to break this degeneracy, demonstrating the limitations of conventional background parameterizations in describing nonlinear DE dynamics. The precision of forthcoming cosmological surveys may provide the sensitivity needed to distinguish between otherwise degenerate DE models through their nonlinear signatures, thereby placing tighter constraints on the space of viable DE theories. Although the analysis here is restricted to spherical collapse, the Lagrangian formulation of the new fluid framework provides a foundation for its implementation in three-dimensional cosmological simulations, with which the observational signatures of this nonlinear degeneracy breaking can be investigated quantitatively. This will be an important next step towards connecting nonlinear, theory-specific, DE dynamics with cosmological observations.

\textbf{Acknowledgements - } YG and JW are supported by the National Key R\&D Program of China (No. 2022YFA1602901). YG acknowledges support from the China Scholarship Council (No.~202510740017). CMB and BL acknowledge support from the UK Science and Technology Facilities Council (STFC) Grant No.~ST/X001075/1. YG and BL are supported by the ERC Advanced Grant UNCA through the UKRI Frontiers Research Guarantee (No.~EP/Z533877/1). JW acknowledges support from the NSFC Grant (No.~12125302), and the CAS Project for Young Scientists in Basic Research Grant (No.~YSBR-062).

\bibliographystyle{apsrev4-2}
\bibliography{reference}

\begin{thebibliography}{25}%
\makeatletter
\providecommand \@ifxundefined [1]{%
 \@ifx{#1\undefined}
}%
\providecommand \@ifnum [1]{%
 \ifnum #1\expandafter \@firstoftwo
 \else \expandafter \@secondoftwo
 \fi
}%
\providecommand \@ifx [1]{%
 \ifx #1\expandafter \@firstoftwo
 \else \expandafter \@secondoftwo
 \fi
}%
\providecommand \natexlab [1]{#1}%
\providecommand \enquote  [1]{``#1''}%
\providecommand \bibnamefont  [1]{#1}%
\providecommand \bibfnamefont [1]{#1}%
\providecommand \citenamefont [1]{#1}%
\providecommand \href@noop [0]{\@secondoftwo}%
\providecommand \href [0]{\begingroup \@sanitize@url \@href}%
\providecommand \@href[1]{\@@startlink{#1}\@@href}%
\providecommand \@@href[1]{\endgroup#1\@@endlink}%
\providecommand \@sanitize@url [0]{\catcode `\\12\catcode `\$12\catcode
  `\&12\catcode `\#12\catcode `\^12\catcode `\_12\catcode `\%12\relax}%
\providecommand \@@startlink[1]{}%
\providecommand \@@endlink[0]{}%
\providecommand \url  [0]{\begingroup\@sanitize@url \@url }%
\providecommand \@url [1]{\endgroup\@href {#1}{\urlprefix }}%
\providecommand \urlprefix  [0]{URL }%
\providecommand \Eprint [0]{\href }%
\providecommand \doibase [0]{https://doi.org/}%
\providecommand \selectlanguage [0]{\@gobble}%
\providecommand \bibinfo  [0]{\@secondoftwo}%
\providecommand \bibfield  [0]{\@secondoftwo}%
\providecommand \translation [1]{[#1]}%
\providecommand \BibitemOpen [0]{}%
\providecommand \bibitemStop [0]{}%
\providecommand \bibitemNoStop [0]{.\EOS\space}%
\providecommand \EOS [0]{\spacefactor3000\relax}%
\providecommand \BibitemShut  [1]{\csname bibitem#1\endcsname}%
\let\auto@bib@innerbib\@empty
\bibitem [{\citenamefont {{Kennedy}}\ \emph {et~al.}(2019)\citenamefont
  {{Kennedy}}, \citenamefont {{Lombriser}},\ and\ \citenamefont
  {{Taylor}}}]{Degeneracy}%
  \BibitemOpen
  \bibfield  {author} {\bibinfo {author} {\bibfnamefont {J.}~\bibnamefont
  {{Kennedy}}}, \bibinfo {author} {\bibfnamefont {L.}~\bibnamefont
  {{Lombriser}}},\ and\ \bibinfo {author} {\bibfnamefont {A.}~\bibnamefont
  {{Taylor}}},\ }\href {https://doi.org/10.1103/PhysRevD.100.044034} {\bibfield
   {journal} {\bibinfo  {journal} {\prd}\ }\textbf {\bibinfo {volume} {100}},\
  \bibinfo {eid} {044034} (\bibinfo {year} {2019})},\ \Eprint
  {https://arxiv.org/abs/1902.09853} {arXiv:1902.09853 [gr-qc]} \BibitemShut
  {NoStop}%
\bibitem [{\citenamefont {{Abdul Karim}}\ \emph {et~al.}(2025)\citenamefont
  {{Abdul Karim}}, \citenamefont {{Aguilar}}, \citenamefont {{Ahlen}} \emph
  {et~al.}}]{DESI01}%
  \BibitemOpen
  \bibfield  {author} {\bibinfo {author} {\bibfnamefont {M.}~\bibnamefont
  {{Abdul Karim}}}, \bibinfo {author} {\bibfnamefont {J.}~\bibnamefont
  {{Aguilar}}}, \bibinfo {author} {\bibfnamefont {S.}~\bibnamefont {{Ahlen}}},
  \emph {et~al.} (\bibinfo {collaboration} {DESI Collaboration}),\ }\href
  {https://doi.org/10.1103/tr6y-kpc6} {\bibfield  {journal} {\bibinfo
  {journal} {\prd}\ }\textbf {\bibinfo {volume} {112}},\ \bibinfo {eid}
  {083515} (\bibinfo {year} {2025})},\ \Eprint
  {https://arxiv.org/abs/2503.14738} {arXiv:2503.14738 [astro-ph.CO]}
  \BibitemShut {NoStop}%
\bibitem [{\citenamefont {{Lodha}}\ \emph {et~al.}(2025)\citenamefont
  {{Lodha}}, \citenamefont {{Calderon}}, \citenamefont {{Matthewson}},
  \citenamefont {{Shafieloo}}, \citenamefont {{Ishak}}, \citenamefont {{Pan}},
  \citenamefont {{Garcia-Quintero}}, \citenamefont {{Huterer}}, \citenamefont
  {{Valogiannis}}, \citenamefont {{Ure{\~n}a-L{\'o}pez}}, \citenamefont
  {{Kamble}}, \citenamefont {{Parkinson}}, \citenamefont {{Kim}}, \citenamefont
  {{Zhao}}, \citenamefont {{Cervantes-Cota}}, \citenamefont {{Rohlf}},
  \citenamefont {{Lozano-Rodr{\'\i}guez}}, \citenamefont {{Rom{\'a}n-Herrera}}
  \emph {et~al.}}]{DESI02}%
  \BibitemOpen
  \bibfield  {author} {\bibinfo {author} {\bibfnamefont {K.}~\bibnamefont
  {{Lodha}}}, \bibinfo {author} {\bibfnamefont {R.}~\bibnamefont {{Calderon}}},
  \bibinfo {author} {\bibfnamefont {W.~L.}\ \bibnamefont {{Matthewson}}},
  \bibinfo {author} {\bibfnamefont {A.}~\bibnamefont {{Shafieloo}}}, \bibinfo
  {author} {\bibfnamefont {M.}~\bibnamefont {{Ishak}}}, \bibinfo {author}
  {\bibfnamefont {J.}~\bibnamefont {{Pan}}}, \bibinfo {author} {\bibfnamefont
  {C.}~\bibnamefont {{Garcia-Quintero}}}, \bibinfo {author} {\bibfnamefont
  {D.}~\bibnamefont {{Huterer}}}, \bibinfo {author} {\bibfnamefont
  {G.}~\bibnamefont {{Valogiannis}}}, \bibinfo {author} {\bibfnamefont {L.~A.}\
  \bibnamefont {{Ure{\~n}a-L{\'o}pez}}}, \bibinfo {author} {\bibfnamefont
  {N.~V.}\ \bibnamefont {{Kamble}}}, \bibinfo {author} {\bibfnamefont
  {D.}~\bibnamefont {{Parkinson}}}, \bibinfo {author} {\bibfnamefont {A.~G.}\
  \bibnamefont {{Kim}}}, \bibinfo {author} {\bibfnamefont {G.~B.}\ \bibnamefont
  {{Zhao}}}, \bibinfo {author} {\bibfnamefont {J.~L.}\ \bibnamefont
  {{Cervantes-Cota}}}, \bibinfo {author} {\bibfnamefont {J.}~\bibnamefont
  {{Rohlf}}}, \bibinfo {author} {\bibfnamefont {F.}~\bibnamefont
  {{Lozano-Rodr{\'\i}guez}}}, \bibinfo {author} {\bibfnamefont {J.~O.}\
  \bibnamefont {{Rom{\'a}n-Herrera}}}, \emph {et~al.} (\bibinfo {collaboration}
  {DESI Collaboration}),\ }\href {https://doi.org/10.1103/w4c6-1r5j} {\bibfield
   {journal} {\bibinfo  {journal} {\prd}\ }\textbf {\bibinfo {volume} {112}},\
  \bibinfo {eid} {083511} (\bibinfo {year} {2025})},\ \Eprint
  {https://arxiv.org/abs/2503.14743} {arXiv:2503.14743 [astro-ph.CO]}
  \BibitemShut {NoStop}%
\bibitem [{\citenamefont {{Gu}}\ \emph {et~al.}(2025)\citenamefont {{Gu}},
  \citenamefont {{Wang}}, \citenamefont {{Wang}}, \citenamefont {{Zhao}},
  \citenamefont {{Pogosian}}, \citenamefont {{Koyama}}, \citenamefont
  {{Peacock}}, \citenamefont {{Cai}}, \citenamefont {{Cervantes-Cota}},
  \citenamefont {{Ishak}}, \citenamefont {{Shafieloo}}, \citenamefont {{Zhao}}
  \emph {et~al.}}]{DESI03}%
  \BibitemOpen
  \bibfield  {author} {\bibinfo {author} {\bibfnamefont {G.}~\bibnamefont
  {{Gu}}}, \bibinfo {author} {\bibfnamefont {X.}~\bibnamefont {{Wang}}},
  \bibinfo {author} {\bibfnamefont {Y.}~\bibnamefont {{Wang}}}, \bibinfo
  {author} {\bibfnamefont {G.-B.}\ \bibnamefont {{Zhao}}}, \bibinfo {author}
  {\bibfnamefont {L.}~\bibnamefont {{Pogosian}}}, \bibinfo {author}
  {\bibfnamefont {K.}~\bibnamefont {{Koyama}}}, \bibinfo {author}
  {\bibfnamefont {J.~A.}\ \bibnamefont {{Peacock}}}, \bibinfo {author}
  {\bibfnamefont {Z.}~\bibnamefont {{Cai}}}, \bibinfo {author} {\bibfnamefont
  {J.~L.}\ \bibnamefont {{Cervantes-Cota}}}, \bibinfo {author} {\bibfnamefont
  {M.}~\bibnamefont {{Ishak}}}, \bibinfo {author} {\bibfnamefont
  {A.}~\bibnamefont {{Shafieloo}}}, \bibinfo {author} {\bibfnamefont
  {R.}~\bibnamefont {{Zhao}}}, \emph {et~al.} (\bibinfo {collaboration} {DESI
  Collaboration}),\ }\href {https://doi.org/10.1038/s41550-025-02669-6}
  {\bibfield  {journal} {\bibinfo  {journal} {Nature Astronomy}\ }\textbf
  {\bibinfo {volume} {9}},\ \bibinfo {pages} {1879} (\bibinfo {year} {2025})},\
  \Eprint {https://arxiv.org/abs/2504.06118} {arXiv:2504.06118 [astro-ph.CO]}
  \BibitemShut {NoStop}%
\bibitem [{\citenamefont {{Mellier}}\ \emph {et~al.}(2025)\citenamefont
  {{Mellier}} \emph {et~al.}}]{Euclid01}%
  \BibitemOpen
  \bibfield  {author} {\bibinfo {author} {\bibfnamefont {Y.}~\bibnamefont
  {{Mellier}}} \emph {et~al.} (\bibinfo {collaboration} {Euclid
  Collaboration}),\ }\href {https://doi.org/10.1051/0004-6361/202450810}
  {\bibfield  {journal} {\bibinfo  {journal} {\aap}\ }\textbf {\bibinfo
  {volume} {697}},\ \bibinfo {eid} {A1} (\bibinfo {year} {2025})},\ \Eprint
  {https://arxiv.org/abs/2405.13491} {arXiv:2405.13491 [astro-ph.CO]}
  \BibitemShut {NoStop}%
\bibitem [{\citenamefont {{Adamek}}\ \emph {et~al.}(2025)\citenamefont
  {{Adamek}}, \citenamefont {{Fiorini}}, \citenamefont {{Baldi}}, \citenamefont
  {{Brando}}, \citenamefont {{Breton}}, \citenamefont {{Hassani}},
  \citenamefont {{Koyama}}, \citenamefont {{Le Brun}}, \citenamefont
  {{R{\'a}cz}}, \citenamefont {{Winther}}, \citenamefont {{Casalino}},
  \citenamefont {{Hern{\'a}ndez-Aguayo}}, \citenamefont {{Li}}, \citenamefont
  {{Potter}}, \citenamefont {{Altamura}}, \citenamefont {{Carbone}},
  \citenamefont {{Giocoli}}, \citenamefont {{Mota}}, \citenamefont
  {{Pourtsidou}}, \citenamefont {{Sakr}}, \citenamefont {{Vernizzi}} \emph
  {et~al.}}]{Euclid02}%
  \BibitemOpen
  \bibfield  {author} {\bibinfo {author} {\bibfnamefont {J.}~\bibnamefont
  {{Adamek}}}, \bibinfo {author} {\bibfnamefont {B.}~\bibnamefont {{Fiorini}}},
  \bibinfo {author} {\bibfnamefont {M.}~\bibnamefont {{Baldi}}}, \bibinfo
  {author} {\bibfnamefont {G.}~\bibnamefont {{Brando}}}, \bibinfo {author}
  {\bibfnamefont {M.-A.}\ \bibnamefont {{Breton}}}, \bibinfo {author}
  {\bibfnamefont {F.}~\bibnamefont {{Hassani}}}, \bibinfo {author}
  {\bibfnamefont {K.}~\bibnamefont {{Koyama}}}, \bibinfo {author}
  {\bibfnamefont {A.~M.~C.}\ \bibnamefont {{Le Brun}}}, \bibinfo {author}
  {\bibfnamefont {G.}~\bibnamefont {{R{\'a}cz}}}, \bibinfo {author}
  {\bibfnamefont {H.-A.}\ \bibnamefont {{Winther}}}, \bibinfo {author}
  {\bibfnamefont {A.}~\bibnamefont {{Casalino}}}, \bibinfo {author}
  {\bibfnamefont {C.}~\bibnamefont {{Hern{\'a}ndez-Aguayo}}}, \bibinfo {author}
  {\bibfnamefont {B.}~\bibnamefont {{Li}}}, \bibinfo {author} {\bibfnamefont
  {D.}~\bibnamefont {{Potter}}}, \bibinfo {author} {\bibfnamefont
  {E.}~\bibnamefont {{Altamura}}}, \bibinfo {author} {\bibfnamefont
  {C.}~\bibnamefont {{Carbone}}}, \bibinfo {author} {\bibfnamefont
  {C.}~\bibnamefont {{Giocoli}}}, \bibinfo {author} {\bibfnamefont {D.~F.}\
  \bibnamefont {{Mota}}}, \bibinfo {author} {\bibfnamefont {A.}~\bibnamefont
  {{Pourtsidou}}}, \bibinfo {author} {\bibfnamefont {Z.}~\bibnamefont
  {{Sakr}}}, \bibinfo {author} {\bibfnamefont {F.}~\bibnamefont {{Vernizzi}}},
  \emph {et~al.} (\bibinfo {collaboration} {Euclid Collaboration}),\ }\href
  {https://doi.org/10.1051/0004-6361/202452180} {\bibfield  {journal} {\bibinfo
   {journal} {\aap}\ }\textbf {\bibinfo {volume} {695}},\ \bibinfo {eid} {A230}
  (\bibinfo {year} {2025})},\ \Eprint {https://arxiv.org/abs/2409.03522}
  {arXiv:2409.03522 [astro-ph.CO]} \BibitemShut {NoStop}%
\bibitem [{\citenamefont {{Frusciante}}\ \emph {et~al.}(2025)\citenamefont
  {{Frusciante}}, \citenamefont {{Martinelli}}, \citenamefont {{Lombriser}},
  \citenamefont {{Silvestri}} \emph {et~al.}}]{Euclid03}%
  \BibitemOpen
  \bibfield  {author} {\bibinfo {author} {\bibfnamefont {N.}~\bibnamefont
  {{Frusciante}}}, \bibinfo {author} {\bibfnamefont {M.}~\bibnamefont
  {{Martinelli}}}, \bibinfo {author} {\bibfnamefont {L.}~\bibnamefont
  {{Lombriser}}}, \bibinfo {author} {\bibfnamefont {A.}~\bibnamefont
  {{Silvestri}}}, \emph {et~al.} (\bibinfo {collaboration} {Euclid
  Collaboration}),\ }\href {https://doi.org/10.48550/arXiv.2512.09748}
  {\bibfield  {journal} {\bibinfo  {journal} {arXiv e-prints}\ ,\ \bibinfo
  {eid} {arXiv:2512.09748}} (\bibinfo {year} {2025})},\ \Eprint
  {https://arxiv.org/abs/2512.09748} {arXiv:2512.09748 [astro-ph.CO]}
  \BibitemShut {NoStop}%
\bibitem [{\citenamefont {{Mandelbaum}}\ \emph {et~al.}(2018)\citenamefont
  {{Mandelbaum}}, \citenamefont {{Eifler}}, \citenamefont {{Hlo{\v{z}}ek}},
  \citenamefont {{Collett}}, \citenamefont {{Gawiser}}, \citenamefont
  {{Scolnic}} \emph {et~al.}}]{LSST01}%
  \BibitemOpen
  \bibfield  {author} {\bibinfo {author} {\bibfnamefont {R.}~\bibnamefont
  {{Mandelbaum}}}, \bibinfo {author} {\bibfnamefont {T.}~\bibnamefont
  {{Eifler}}}, \bibinfo {author} {\bibfnamefont {R.}~\bibnamefont
  {{Hlo{\v{z}}ek}}}, \bibinfo {author} {\bibfnamefont {T.}~\bibnamefont
  {{Collett}}}, \bibinfo {author} {\bibfnamefont {E.}~\bibnamefont
  {{Gawiser}}}, \bibinfo {author} {\bibfnamefont {D.}~\bibnamefont
  {{Scolnic}}}, \emph {et~al.} (\bibinfo {collaboration} {The LSST Dark Energy
  Science Collaboration}),\ }\href {https://doi.org/10.48550/arXiv.1809.01669}
  {\bibfield  {journal} {\bibinfo  {journal} {arXiv e-prints}\ ,\ \bibinfo
  {eid} {arXiv:1809.01669}} (\bibinfo {year} {2018})},\ \Eprint
  {https://arxiv.org/abs/1809.01669} {arXiv:1809.01669 [astro-ph.CO]}
  \BibitemShut {NoStop}%
\bibitem [{\citenamefont {{Ivezi{\'c}}}\ \emph {et~al.}(2019)\citenamefont
  {{Ivezi{\'c}}}, \citenamefont {{Kahn}}, \citenamefont {{Tyson}} \emph
  {et~al.}}]{LSST02}%
  \BibitemOpen
  \bibfield  {author} {\bibinfo {author} {\bibfnamefont {{\v{Z}}.}~\bibnamefont
  {{Ivezi{\'c}}}}, \bibinfo {author} {\bibfnamefont {S.~M.}\ \bibnamefont
  {{Kahn}}}, \bibinfo {author} {\bibfnamefont {J.~A.}\ \bibnamefont {{Tyson}}},
  \emph {et~al.},\ }\href {https://doi.org/10.3847/1538-4357/ab042c} {\bibfield
   {journal} {\bibinfo  {journal} {\apj}\ }\textbf {\bibinfo {volume} {873}},\
  \bibinfo {eid} {111} (\bibinfo {year} {2019})},\ \Eprint
  {https://arxiv.org/abs/0805.2366} {arXiv:0805.2366 [astro-ph]} \BibitemShut
  {NoStop}%
\bibitem [{\citenamefont {{Mota}}\ and\ \citenamefont {{van de
  Bruck}}(2004)}]{SC01}%
  \BibitemOpen
  \bibfield  {author} {\bibinfo {author} {\bibfnamefont {D.~F.}\ \bibnamefont
  {{Mota}}}\ and\ \bibinfo {author} {\bibfnamefont {C.}~\bibnamefont {{van de
  Bruck}}},\ }\href {https://doi.org/10.1051/0004-6361:20041090} {\bibfield
  {journal} {\bibinfo  {journal} {\aap}\ }\textbf {\bibinfo {volume} {421}},\
  \bibinfo {pages} {71} (\bibinfo {year} {2004})},\ \Eprint
  {https://arxiv.org/abs/astro-ph/0401504} {arXiv:astro-ph/0401504 [astro-ph]}
  \BibitemShut {NoStop}%
\bibitem [{\citenamefont {{Abramo}}\ \emph {et~al.}(2007)\citenamefont
  {{Abramo}}, \citenamefont {{Batista}}, \citenamefont {{Liberato}},\ and\
  \citenamefont {{Rosenfeld}}}]{SC02}%
  \BibitemOpen
  \bibfield  {author} {\bibinfo {author} {\bibfnamefont {L.~R.}\ \bibnamefont
  {{Abramo}}}, \bibinfo {author} {\bibfnamefont {R.~C.}\ \bibnamefont
  {{Batista}}}, \bibinfo {author} {\bibfnamefont {L.}~\bibnamefont
  {{Liberato}}},\ and\ \bibinfo {author} {\bibfnamefont {R.}~\bibnamefont
  {{Rosenfeld}}},\ }\href {https://doi.org/10.1088/1475-7516/2007/11/012}
  {\bibfield  {journal} {\bibinfo  {journal} {\jcap}\ }\textbf {\bibinfo
  {volume} {2007}},\ \bibinfo {eid} {012} (\bibinfo {year} {2007})},\ \Eprint
  {https://arxiv.org/abs/0707.2882} {arXiv:0707.2882 [astro-ph]} \BibitemShut
  {NoStop}%
\bibitem [{\citenamefont {{Creminelli}}\ \emph {et~al.}(2010)\citenamefont
  {{Creminelli}}, \citenamefont {{D'Amico}}, \citenamefont {{Nore{\~n}a}},
  \citenamefont {{Senatore}},\ and\ \citenamefont {{Vernizzi}}}]{SC03}%
  \BibitemOpen
  \bibfield  {author} {\bibinfo {author} {\bibfnamefont {P.}~\bibnamefont
  {{Creminelli}}}, \bibinfo {author} {\bibfnamefont {G.}~\bibnamefont
  {{D'Amico}}}, \bibinfo {author} {\bibfnamefont {J.}~\bibnamefont
  {{Nore{\~n}a}}}, \bibinfo {author} {\bibfnamefont {L.}~\bibnamefont
  {{Senatore}}},\ and\ \bibinfo {author} {\bibfnamefont {F.}~\bibnamefont
  {{Vernizzi}}},\ }\href {https://doi.org/10.1088/1475-7516/2010/03/027}
  {\bibfield  {journal} {\bibinfo  {journal} {\jcap}\ }\textbf {\bibinfo
  {volume} {2010}},\ \bibinfo {eid} {027} (\bibinfo {year} {2010})},\ \Eprint
  {https://arxiv.org/abs/0911.2701} {arXiv:0911.2701 [astro-ph.CO]}
  \BibitemShut {NoStop}%
\bibitem [{\citenamefont {{Basse}}\ \emph {et~al.}(2011)\citenamefont
  {{Basse}}, \citenamefont {{Eggers Bj{\ae}lde}},\ and\ \citenamefont
  {{Wong}}}]{SC04}%
  \BibitemOpen
  \bibfield  {author} {\bibinfo {author} {\bibfnamefont {T.}~\bibnamefont
  {{Basse}}}, \bibinfo {author} {\bibfnamefont {O.}~\bibnamefont {{Eggers
  Bj{\ae}lde}}},\ and\ \bibinfo {author} {\bibfnamefont {Y.~Y.~Y.}\
  \bibnamefont {{Wong}}},\ }\href
  {https://doi.org/10.1088/1475-7516/2011/10/038} {\bibfield  {journal}
  {\bibinfo  {journal} {\jcap}\ }\textbf {\bibinfo {volume} {2011}},\ \bibinfo
  {eid} {038} (\bibinfo {year} {2011})},\ \Eprint
  {https://arxiv.org/abs/1009.0010} {arXiv:1009.0010 [astro-ph.CO]}
  \BibitemShut {NoStop}%
\bibitem [{\citenamefont {{Wang}}\ and\ \citenamefont {{Fan}}(2012)}]{SC05}%
  \BibitemOpen
  \bibfield  {author} {\bibinfo {author} {\bibfnamefont {Q.}~\bibnamefont
  {{Wang}}}\ and\ \bibinfo {author} {\bibfnamefont {Z.}~\bibnamefont {{Fan}}},\
  }\href {https://doi.org/10.1103/PhysRevD.85.023002} {\bibfield  {journal}
  {\bibinfo  {journal} {\prd}\ }\textbf {\bibinfo {volume} {85}},\ \bibinfo
  {eid} {023002} (\bibinfo {year} {2012})},\ \Eprint
  {https://arxiv.org/abs/1112.4268} {arXiv:1112.4268 [astro-ph.CO]}
  \BibitemShut {NoStop}%
\bibitem [{\citenamefont {{Batista}}\ \emph {et~al.}(2023)\citenamefont
  {{Batista}}, \citenamefont {{de Oliveira}},\ and\ \citenamefont
  {{Abramo}}}]{SC06}%
  \BibitemOpen
  \bibfield  {author} {\bibinfo {author} {\bibfnamefont {R.~C.}\ \bibnamefont
  {{Batista}}}, \bibinfo {author} {\bibfnamefont {H.~P.}\ \bibnamefont {{de
  Oliveira}}},\ and\ \bibinfo {author} {\bibfnamefont {L.~R.~W.}\ \bibnamefont
  {{Abramo}}},\ }\href {https://doi.org/10.1088/1475-7516/2023/02/037}
  {\bibfield  {journal} {\bibinfo  {journal} {\jcap}\ }\textbf {\bibinfo
  {volume} {2023}},\ \bibinfo {eid} {037} (\bibinfo {year} {2023})},\ \Eprint
  {https://arxiv.org/abs/2210.14769} {arXiv:2210.14769 [astro-ph.CO]}
  \BibitemShut {NoStop}%
\bibitem [{\citenamefont {{Sefusatti}}\ and\ \citenamefont
  {{Vernizzi}}(2011)}]{Perturbation_theory01}%
  \BibitemOpen
  \bibfield  {author} {\bibinfo {author} {\bibfnamefont {E.}~\bibnamefont
  {{Sefusatti}}}\ and\ \bibinfo {author} {\bibfnamefont {F.}~\bibnamefont
  {{Vernizzi}}},\ }\href {https://doi.org/10.1088/1475-7516/2011/03/047}
  {\bibfield  {journal} {\bibinfo  {journal} {\jcap}\ }\textbf {\bibinfo
  {volume} {2011}},\ \bibinfo {eid} {047} (\bibinfo {year} {2011})},\ \Eprint
  {https://arxiv.org/abs/1101.1026} {arXiv:1101.1026 [astro-ph.CO]}
  \BibitemShut {NoStop}%
\bibitem [{\citenamefont {{Anselmi}}\ \emph {et~al.}(2011)\citenamefont
  {{Anselmi}}, \citenamefont {{Ballesteros}},\ and\ \citenamefont
  {{Pietroni}}}]{Perturbation_theory02}%
  \BibitemOpen
  \bibfield  {author} {\bibinfo {author} {\bibfnamefont {S.}~\bibnamefont
  {{Anselmi}}}, \bibinfo {author} {\bibfnamefont {G.}~\bibnamefont
  {{Ballesteros}}},\ and\ \bibinfo {author} {\bibfnamefont {M.}~\bibnamefont
  {{Pietroni}}},\ }\href {https://doi.org/10.1088/1475-7516/2011/11/014}
  {\bibfield  {journal} {\bibinfo  {journal} {\jcap}\ }\textbf {\bibinfo
  {volume} {2011}},\ \bibinfo {eid} {014} (\bibinfo {year} {2011})},\ \Eprint
  {https://arxiv.org/abs/1106.0834} {arXiv:1106.0834 [astro-ph.CO]}
  \BibitemShut {NoStop}%
\bibitem [{\citenamefont {{Anselmi}}\ \emph {et~al.}(2014)\citenamefont
  {{Anselmi}}, \citenamefont {{L{\'o}pez Nacir}},\ and\ \citenamefont
  {{Sefusatti}}}]{Perturbation_theory03}%
  \BibitemOpen
  \bibfield  {author} {\bibinfo {author} {\bibfnamefont {S.}~\bibnamefont
  {{Anselmi}}}, \bibinfo {author} {\bibfnamefont {D.}~\bibnamefont {{L{\'o}pez
  Nacir}}},\ and\ \bibinfo {author} {\bibfnamefont {E.}~\bibnamefont
  {{Sefusatti}}},\ }\href {https://doi.org/10.1088/1475-7516/2014/07/013}
  {\bibfield  {journal} {\bibinfo  {journal} {\jcap}\ }\textbf {\bibinfo
  {volume} {2014}},\ \bibinfo {eid} {013} (\bibinfo {year} {2014})},\ \Eprint
  {https://arxiv.org/abs/1402.4269} {arXiv:1402.4269 [astro-ph.CO]}
  \BibitemShut {NoStop}%
\bibitem [{\citenamefont {{Hassani}}\ \emph {et~al.}(2019)\citenamefont
  {{Hassani}}, \citenamefont {{Adamek}}, \citenamefont {{Kunz}},\ and\
  \citenamefont {{Vernizzi}}}]{k-evolution01}%
  \BibitemOpen
  \bibfield  {author} {\bibinfo {author} {\bibfnamefont {F.}~\bibnamefont
  {{Hassani}}}, \bibinfo {author} {\bibfnamefont {J.}~\bibnamefont {{Adamek}}},
  \bibinfo {author} {\bibfnamefont {M.}~\bibnamefont {{Kunz}}},\ and\ \bibinfo
  {author} {\bibfnamefont {F.}~\bibnamefont {{Vernizzi}}},\ }\href
  {https://doi.org/10.1088/1475-7516/2019/12/011} {\bibfield  {journal}
  {\bibinfo  {journal} {\jcap}\ }\textbf {\bibinfo {volume} {2019}},\ \bibinfo
  {eid} {011} (\bibinfo {year} {2019})},\ \Eprint
  {https://arxiv.org/abs/1910.01104} {arXiv:1910.01104 [astro-ph.CO]}
  \BibitemShut {NoStop}%
\bibitem [{\citenamefont {{Nouri-Zonoz}}\ \emph {et~al.}(2026)\citenamefont
  {{Nouri-Zonoz}}, \citenamefont {{Hassani}}, \citenamefont {{Bellini}},\ and\
  \citenamefont {{Kunz}}}]{k-evolution02}%
  \BibitemOpen
  \bibfield  {author} {\bibinfo {author} {\bibfnamefont {A.}~\bibnamefont
  {{Nouri-Zonoz}}}, \bibinfo {author} {\bibfnamefont {F.}~\bibnamefont
  {{Hassani}}}, \bibinfo {author} {\bibfnamefont {E.}~\bibnamefont
  {{Bellini}}},\ and\ \bibinfo {author} {\bibfnamefont {M.}~\bibnamefont
  {{Kunz}}},\ }\href {https://doi.org/10.1088/1475-7516/2026/08/011} {\bibfield
   {journal} {\bibinfo  {journal} {\jcap}\ }\textbf {\bibinfo {volume}
  {2026}},\ \bibinfo {eid} {011} (\bibinfo {year} {2026})},\ \Eprint
  {https://arxiv.org/abs/2511.04676} {arXiv:2511.04676 [astro-ph.CO]}
  \BibitemShut {NoStop}%
\bibitem [{\citenamefont {{Blot}}\ \emph {et~al.}(2023)\citenamefont {{Blot}},
  \citenamefont {{Corasaniti}},\ and\ \citenamefont {{Schmidt}}}]{FVM01}%
  \BibitemOpen
  \bibfield  {author} {\bibinfo {author} {\bibfnamefont {L.}~\bibnamefont
  {{Blot}}}, \bibinfo {author} {\bibfnamefont {P.~S.}\ \bibnamefont
  {{Corasaniti}}},\ and\ \bibinfo {author} {\bibfnamefont {F.}~\bibnamefont
  {{Schmidt}}},\ }\href {https://doi.org/10.1088/1475-7516/2023/05/001}
  {\bibfield  {journal} {\bibinfo  {journal} {\jcap}\ }\textbf {\bibinfo
  {volume} {2023}},\ \bibinfo {eid} {001} (\bibinfo {year} {2023})},\ \Eprint
  {https://arxiv.org/abs/2210.04800} {arXiv:2210.04800 [astro-ph.CO]}
  \BibitemShut {NoStop}%
\bibitem [{\citenamefont {{Blot}}\ \emph {et~al.}(2026)\citenamefont {{Blot}},
  \citenamefont {{Gayoux}}, \citenamefont {{Schmidt}}, \citenamefont
  {{Corasaniti}},\ and\ \citenamefont {{de Ligondes}}}]{FVM02}%
  \BibitemOpen
  \bibfield  {author} {\bibinfo {author} {\bibfnamefont {L.}~\bibnamefont
  {{Blot}}}, \bibinfo {author} {\bibfnamefont {T.}~\bibnamefont {{Gayoux}}},
  \bibinfo {author} {\bibfnamefont {F.}~\bibnamefont {{Schmidt}}}, \bibinfo
  {author} {\bibfnamefont {P.~S.}\ \bibnamefont {{Corasaniti}}},\ and\ \bibinfo
  {author} {\bibfnamefont {B.}~\bibnamefont {{de Ligondes}}},\ }\href
  {https://doi.org/10.48550/arXiv.2603.16248} {\bibfield  {journal} {\bibinfo
  {journal} {arXiv e-prints}\ ,\ \bibinfo {eid} {arXiv:2603.16248}} (\bibinfo
  {year} {2026})},\ \Eprint {https://arxiv.org/abs/2603.16248}
  {arXiv:2603.16248 [astro-ph.CO]} \BibitemShut {NoStop}%
\bibitem [{\citenamefont {{Hassani}}\ \emph {et~al.}(2021)\citenamefont
  {{Hassani}}, \citenamefont {{Adamek}},\ and\ \citenamefont
  {{Kunz}}}]{k-evolution-observation}%
  \BibitemOpen
  \bibfield  {author} {\bibinfo {author} {\bibfnamefont {F.}~\bibnamefont
  {{Hassani}}}, \bibinfo {author} {\bibfnamefont {J.}~\bibnamefont
  {{Adamek}}},\ and\ \bibinfo {author} {\bibfnamefont {M.}~\bibnamefont
  {{Kunz}}},\ }\href {https://doi.org/10.1093/mnras/staa3589} {\bibfield
  {journal} {\bibinfo  {journal} {\mnras}\ }\textbf {\bibinfo {volume} {500}},\
  \bibinfo {pages} {4514} (\bibinfo {year} {2021})},\ \Eprint
  {https://arxiv.org/abs/2007.04968} {arXiv:2007.04968 [astro-ph.CO]}
  \BibitemShut {NoStop}%
\bibitem [{\citenamefont {Gao}\ \emph {et~al.}(2026)\citenamefont {Gao},
  \citenamefont {Li}, \citenamefont {Wang},\ and\ \citenamefont
  {Baugh}}]{CompanionPRD}%
  \BibitemOpen
  \bibfield  {author} {\bibinfo {author} {\bibfnamefont {Y.}~\bibnamefont
  {Gao}}, \bibinfo {author} {\bibfnamefont {B.}~\bibnamefont {Li}}, \bibinfo
  {author} {\bibfnamefont {J.}~\bibnamefont {Wang}},\ and\ \bibinfo {author}
  {\bibfnamefont {C.}~\bibnamefont {Baugh}},\ }\href
  {https://doi.org/10.48550/arXiv.2611.26461} {\bibinfo {title} {Splitting the
  action: A fully nonlinear fluid method for clustering dark energy}} (\bibinfo
  {year} {2026}),\ \bibinfo {note} {companion paper},\ \Eprint
  {https://arxiv.org/abs/2611.26461} {arXiv:2611.26461 [astro-ph.CO]}
  \BibitemShut {NoStop}%
\bibitem [{\citenamefont {{Gao}}\ \emph {et~al.}(2025)\citenamefont {{Gao}},
  \citenamefont {{Li}},\ and\ \citenamefont {{Wang}}}]{Reconstruction}%
  \BibitemOpen
  \bibfield  {author} {\bibinfo {author} {\bibfnamefont {Y.}~\bibnamefont
  {{Gao}}}, \bibinfo {author} {\bibfnamefont {B.}~\bibnamefont {{Li}}},\ and\
  \bibinfo {author} {\bibfnamefont {J.}~\bibnamefont {{Wang}}},\ }\href
  {https://doi.org/10.1088/1475-7516/2025/10/057} {\bibfield  {journal}
  {\bibinfo  {journal} {\jcap}\ }\textbf {\bibinfo {volume} {2025}},\ \bibinfo
  {eid} {057} (\bibinfo {year} {2025})},\ \Eprint
  {https://arxiv.org/abs/2507.01442} {arXiv:2507.01442 [astro-ph.CO]}
  \BibitemShut {NoStop}%
\end{thebibliography}%

\appendix
\section{Perturbations of the Theory-Dependent Quantities}
We first connect the partial derivatives of $K$ to 
fluid quantities. On the background, we have
\begin{equation}\label{K KX KXX}
\begin{aligned}
    K&=P_{\rm DE}(a)=\rho_{\rm DE}w,\\
    XK_X&=\rho_k(a)=\frac{1}{2}\rho_{\rm DE}(1+w),\\
    X^2K_{XX}&=\frac{1}{2}\left(\frac{1}{\bar{c}_s^2(a)}-1\right)XK_X,
\end{aligned}
\end{equation}
and the continuity equation 
\begin{equation}\label{background continuity}
    \dot{\rho}_{\rm DE}+3H\rho_{\rm DE}(1+w)=0,
\end{equation}
where the dot denotes the derivative 
with respect to the coordinate time. Using field redefinition, one can fix $\phi=a$ on the background, and thus $X=-a^2H^2$.\footnote{For the sake of simplicity, we omit the bar above notation when denoting background quantities.} As a result,
\begin{equation}\label{background scalar}
    \frac{\dot{\phi}}{\phi}=H, \;\frac{\dot{X}}{X}=2\left(H+\frac{\dot{H}}{H}\right)=-H\left(1+3w\Omega_{\rm DE}(a)\right),
\end{equation}
where $\Omega_{\rm DE}(a)$ is the density 
parameter of DE at $a$ (for brevity we shall drop the $a$-dependence in the rest of this appendix). Through these relations, as well as
\begin{equation}
    \dot{K}=K_\phi\dot{\phi}+K_X\dot{X},
\end{equation}
we can solve for $\phi K_\phi$ as
\begin{equation}\label{Kphi}
    \phi K_\phi=\frac{1}{2}\rho_{\rm DE}(1+w)\left[\frac{2w'}{1+w}-6w+1+3w\Omega_{\rm DE}\right],
\end{equation}
with $'$ being the derivative 
with respect to $\ln{a}$. Similarly, by computing the time derivatives of $XK_X$ and $X^2K_{XX}$ and substituting Eqs.~\eqref{K KX KXX}~\eqref{background continuity}~\eqref{background scalar}, we can obtain the background formulas of $\phi XK_{\phi X}$ and $\phi X^2K_{\phi XX}$:
\begin{align}\label{KphiX KphiXX}
    \phi XK_{\phi X}=&\;\frac{1}{2}\rho_{\rm DE}(1+w)\times\nonumber\\&\left[\frac{w'}{1+w}-3(1+w)+\frac{1+\bar{c}_s^2}{2\bar{c}_s^2}(1+3w\Omega_{\rm DE})\right],\nonumber\\
    \phi X^2K_{\phi XX}=&\;\frac{1-\bar{c}_s^2}{4\bar{c}_s^2}\rho_{\rm DE}(1+w)\times\nonumber\\
    &\Bigg[\frac{w'}{1+w}-3(1+w)-\frac{(\ln{\bar{c}_s^2})'}{1-\bar{c}_s^2}\nonumber\\
    &+\left(Q+\frac{1}{2}\right)(1+3w\Omega_{\rm DE})\Bigg].
\end{align}
Here, we define $Q=\dfrac{XK_{XXX}}{K_{XX}}+\dfrac{3}{2}$, which involves the third-order derivatives and can not be expressed by normal fluid quantities. 
Using this definition, we have 
\begin{equation}\label{KXXX}
    X^3K_{XXX}=\left(Q-\frac{3}{2}\right)X^2K_{XX}.
\end{equation}

For any non-zero function $g(\phi, X)$, its relative perturbation can be estimated as
\begin{equation}\label{d ln g 1}
    \delta\ln{|g|}=\frac{g_\phi}{g}\delta\phi+\frac{g_X}{g}\delta X.
\end{equation}
Differentiating the 
expression for the DE density $\rho_{\rm DE}=2XK_X-K$ on both sides, we can derive the relation $\text{d}\rho_{\rm DE}=(K_X+2XK_{XX})\text{d}X+(2XK_{X\phi}-K_\phi)\text{d}\phi$. Replacing $\delta X$ in Eq.~\eqref{d ln g 1} through this relation gives 
\begin{equation}
\begin{aligned}
    \delta\ln{|g|}=&\,\frac{2\bar{c}_s^2}{1+w}\frac{\partial\ln{|g|}}{\partial\ln{(-X)}}\delta\ln{\rho_{\rm DE}}
    +\Bigg[\frac{\partial\ln{|g|}}{\partial\ln{|\phi|}} \\&+\bar{c}_s^2\frac{\partial\ln{|g|}}{\partial\ln{(-X)}}\left(\frac{\phi K_\phi}{XK_X}-2\frac{\partial\ln{|\rho_k|}}{\partial\ln{|\phi|}}\right)\Bigg]\delta\ln{|\phi|},
\end{aligned}
\end{equation}
where the pre-factors of $\delta\ln{\rho}_{\rm DE}$ and $\delta\ln{|\phi|}$ are evaluated on the background. Consequently, knowing the order of these pre-factors as well as $\delta\ln{\rho}_{\rm DE}$ and $\delta\ln{|\phi|}$ can help us determine the order of $\delta\ln{|g|}$.

Now, 
for $K$ and $F = XK_{XX}/K_X$, we have
\begin{equation}
\begin{aligned}
    \delta\ln{|K|}=&\;\frac{2\bar{c}_s^2}{1+w}\frac{XK_X}{K}\delta\ln{|\rho_{\rm DE}|}+\Bigg[\frac{\phi K_\phi}{K}\\&+\bar{c}_s^2\frac{XK_X}{K}\left(\frac{\phi K_{\phi}}{XK_{X}}-\frac{2\phi XK_{\phi X}}{XK_X}\right)\Bigg]\delta\ln{|\phi|},\\
    \delta\ln{|F|}=&\;\frac{2\bar{c}_s^2}{1+w}\left[1+\frac{XK_{XXX}}{K_{XX}}-\frac{XK_{XX}}{K_{X}}\right]\delta\ln{|\rho_{\rm DE}|}\\
    &+\Bigg[\frac{\phi K_{\phi XX}}{K_{XX}}+\bar{c}_s^2\left(\frac{\phi K_{\phi}}{XK_{X}}-\frac{2\phi XK_{\phi X}}{XK_X}\right)\times\\&\left(1+\frac{XK_{XXX}}{K_{XX}}-\frac{XK_{XX}}{K_{X}}\right)-\frac{\phi K_{\phi X}}{K_{X}}\Bigg]\delta\ln{|\phi|}.
\end{aligned}
\end{equation}
By substituting Eqs.~\eqref{K KX KXX}~\eqref{Kphi}~\eqref{KphiX KphiXX}~\eqref{KXXX} into these, we can derive Eq.~\eqref{d ln K} and Eq.~\eqref{d ln F}. It is also worth noting that, for the SV model defined in Eq.~\eqref{XE and SV}, we can prove that $F_{\rm SV}$ is a constant, and thus $\dfrac{\partial\ln{F}_{\rm SV}}{\partial\ln{|\phi|}}=\dfrac{\partial\ln{F}_{\rm SV}}{\partial\ln{|X|}}=0$. Consequently, $\delta\ln{F}_{\rm SV}=0$ over the simulation, which explains the radius independent $c_s^2$ profile in Figure~\ref{fig:SoundSpeed_and_Source}.

For $S$, the perturbation can be expressed as
\begin{equation}
\begin{aligned}
    \delta S=&\;\frac{2\bar{c}_s^2}{1+w}XS_X\delta\ln{|\rho_{\rm DE}|}\\
    +&\;\Bigg[\phi S_\phi+\bar{c}_s^2XS_X\left(\frac{\phi K_{\phi}}{XK_{X}}-\frac{2\phi XK_{\phi X}}{XK_X}\right)\Bigg]\delta\ln{|\phi|},
\end{aligned}
\end{equation}
where
\begin{equation}
\begin{aligned}
    S_X=&\;\;\bar{c}^2_{s,X}(-X)^{3/2}K_{\phi X}+\bar{c}_s^2(-X)^{3/2}K_{\phi XX}\\&+\frac{1}{2}\bar{c}^2_{s,X}\sqrt{-X}K_\phi+\frac{1-2\bar{c}_s^2}{2}\sqrt{-X}K_{\phi X}\\&-\frac{1+\bar{c}_s^2}{4}\frac{K_\phi}{\sqrt{-X}},\\
    S_\phi=&\;\bar{c}^2_{s,\phi}(-X)^{3/2}K_{\phi X}+\bar{c}_s^2(-X)^{3/2}K_{\phi\phi X}\\&+\frac{1}{2}\bar{c}^2_{s,\phi}\sqrt{-X}K_\phi+\frac{1+\bar{c}_s^2}{2}\sqrt{-X}K_{\phi\phi}.
\end{aligned}
\end{equation}
We also have
\begin{equation}\label{cs2 X}
\begin{aligned}
    X\bar{c}_{s,X}^2&=\bar{c}_s^2\left[\frac{XK_{XX}}{K_X}-\frac{3XK_{XX}+2X^2K_{XXX}}{K_X+2XK_{XX}}\right]\\
    &=\frac{1-\bar{c}_s^2}{2}(1-2\bar{c}_s^2Q).
\end{aligned}
\end{equation}
Taking this relation along with Eqs.~\eqref{K KX KXX}~\eqref{Kphi}~\eqref{KphiX KphiXX}~\eqref{KXXX}, we can obtain $T_1(a)$ in Eq.~\eqref{d S}. To derive $T_2(a)$, we can resort to the relation
\begin{equation}\label{dS/dlna}
    \frac{\text{d}S}{\text{d}\ln{a}} = \frac{2\bar{c}_s^2}{1+w}H\rho_k T_1(a)\frac{\text{d}\ln{\rho}_{\rm DE}}{\text{d}\ln{a}}+H\rho_k T_2(a)\frac{\text{d}\ln{|\phi|}}{\text{d}\ln{a}}.
\end{equation}
Note that
\begin{equation}
    \frac{\text{d}\ln{\rho}_{\rm DE}}{\text{d}\ln{a}}=-3(1+w),\;\; \frac{\text{d}\ln{|\phi|}}{\text{d}\ln{a}}=1,
\end{equation}
and on the background we also have
\begin{equation}
    S(a)=\frac{1}{2}\rho_{\rm DE}H(1+w)\left[3\bar{c}_s^2+\frac{w'}{1+w}-3w\right].
\end{equation}
Consequently, we can solve for $T_2(a)$ in Eq.~\eqref{d S} from Eq.~\eqref{dS/dlna}.

As a side remark, we note that the radius-dependent features in $S_1=\dfrac{1+c_s^2}{2}\sqrt{-X}K_\phi$ and $S_2=c_s^2(-X)^{3/2}K_{\phi X}$ 
in Figure~\ref{fig:SoundSpeed_and_Source} are caused by some $\mathcal{O}(1)$ terms in their perturbation expansion. For example, for $S_1$, there are terms like
\begin{equation}\label{d S1}
    \delta S_1=H\rho_k\left(\frac{1+3w\Omega_{\rm DE}}{2(1+w)}+\cdots\right)\delta\ln{\rho}_{\rm DE}+\left(\cdots\right)\delta\ln{|\phi|}.
\end{equation}
As $\delta\ln{\rho}_{\rm DE}$ can be of order $\mathcal{O}(1)$ at non-linear scales, the first term on the RHS of the above equation can lead to $\mathcal{O}(1)$ perturbations in the local value of $S_1$. The analysis for $S_2$ yields a similar result. However, when adding $S_1$ and $S_2$ together, the $\mathcal{O}(1)$ perturbation terms cancel out, so that the perturbation of $S$ stays at $\mathcal{O}(\bar{c}_s^2)$ and $\mathcal{O}(\delta\phi/\bar{\phi})$.

\end{document}